\documentclass[
pra,
twocolumn,
amsmath,amsfonts,
superscriptaddress,
]{revtex4-2}
\usepackage{hyperref}
\usepackage{amssymb}
\usepackage{graphicx}
\usepackage{import}
\usepackage{bbm}

\usepackage{tikz}
\usepackage{pgfplots}
\usepgfplotslibrary{fillbetween}
\pgfplotsset{compat=newest}
\usetikzlibrary{plotmarks}
\usetikzlibrary{arrows.meta}
\usepgfplotslibrary{patchplots}
\usepackage{grffile}

\definecolor{mred}{rgb}{0.9, 0.1, 0.29}%
\definecolor{mgreen}{rgb}{0.24, 0.71, 0.29}%
\definecolor{myellow}{rgb}{1, 1, 0.1}%
\definecolor{mblue}{rgb}{0, 0.51, 0.78}%
\definecolor{morange}{rgb}{0.96, 0.51, 0.19}%
\definecolor{mcyan}{rgb}{0.27, 0.94, 0.94}%
\definecolor{mmagenta}{rgb}{0.94, 0.20, 0.90}%
\definecolor{mpink}{rgb}{0.98, 0.75, 0.83}%
\definecolor{mbrown}{rgb}{0.66, 0.43, 0.16}%
\definecolor{mnavy}{rgb}{0.0, 0.0, 0.5}%
\definecolor{mgray}{rgb}{0.5, 0.5, 0.5}%

\newcommand{\tf} {{g}}

\newcommand{\adtol} {{\rm tol \, }}
\newcommand{\SVtol} {{\rm SV tol.}}
\newcommand{\param} {{\theta}}
\renewcommand{\arg} {{x}}
\newcommand{\argLim} {{X}}
\newcommand{\Darg} {{\Delta\arg}}
\newcommand{\im} {{\iota}}
\newcommand{\dT} {\dot}
\newcommand{\tr} {{\rm tr}}
\newcommand{\K} {{\mathcal{K}_\param}}
\newcommand{\sld} {{\mathcal{L}_\param}}
\newcommand{\B} {{\mathcal{B}_\param}}
\newcommand{\A} {{\mathcal{A}_\param}}
\newcommand{\qfi} {{\mathcal{F}_\param}}

\newcommand{\dqfi} {{\bar \qfi}}
\newcommand{\rhoParam}[1] {{\rho}_{#1}}
\newcommand{\relErr} {{\epsilon_\param}}
\newcommand{\ket}[1] {{|{#1}\rangle}}
\newcommand{\bra}[1] {{\langle{#1}|}}

\newcommand{\x}[1] {{\sigma^x_{#1}}}
\newcommand{\z}[1] {{\sigma^z_{#1}}}

\renewcommand{\[}{\begin{equation}}
\renewcommand{\]}{\end{equation}}

\begin{document}

\title{Quantum Fisher information from tensor network integration of Lyapunov equation}
\author{Gabriela Wójtowicz}
\email{gabriela.wojtowicz@uni-ulm.de}
\affiliation{Institut f\"ur Theoretische Physik und IQST, Albert-Einstein-Allee 11, Universit\"at Ulm, D-89081 Ulm, Germany}%
\author{Susana F. Huelga}
\affiliation{Institut f\"ur Theoretische Physik und IQST, Albert-Einstein-Allee 11, Universit\"at Ulm, D-89081 Ulm, Germany}%
\author{Marek M. Rams}
\affiliation{Institute of Theoretical Physics, Jagiellonian University, Łojasiewicza 11, 30-348 Kraków, Poland}
\affiliation{Mark Kac Complex Systems Research Center, Jagiellonian University, Łojasiewicza 11, 30-348 Kraków, Poland}
\author{Martin B. Plenio}
\affiliation{Institut f\"ur Theoretische Physik und IQST, Albert-Einstein-Allee 11, Universit\"at Ulm, D-89081 Ulm, Germany}%

\begin{abstract}
The Quantum Fisher Information (QFI) is a geometric measure of state deformation calculated along the trajectory parameterizing an ensemble of quantum states. 
It serves as a key concept in quantum metrology, where it is linked to the fundamental limit on the precision of the parameter that we estimate. 
However, the QFI is notoriously difficult to calculate due to its non-linear mathematical form. 
For mixed states, standard numerical procedures based on eigendecomposition quickly become impractical with increasing system size. 
To overcome this limitation, we introduce a novel numerical approach based on Lyapunov integrals that combines the concept of symmetric logarithmic derivative and tensor networks.  
Importantly, this approach requires only the elementary matrix product states algorithm for time-evolution, opening a perspective for broad usage and application to many-body systems. 
We discuss the advantages and limitations of our methodology through an illustrative example in quantum metrology, where the thermal state of the transverse-field Ising model is used to measure magnetic field amplitude.
\end{abstract}

\maketitle

Estimating the distinguishability of quantum states through measurements lies at the heart of quantum sciences, including quantum metrology and state discrimination~\cite{uhlmann_metric_1992,braunstein_statistical_1994,paris_quantum_2009,giovannetti_advances_2011,toth_quantum_2014}.
The {\it Quantum Fisher information} (QFI), which is proportional to the Bures metric~\cite{helstrom_minimum_1967,bures_extension_1969}, is a key measure quantifying the distinguishability of quantum states under an infinitesimal parameter change~\cite{braunstein_statistical_1994,braunstein_generalized_1996,paris_quantum_2009}. 
It can be used as a witness to quantum phase transitions~\cite{zanardi_quantum_2008}, to detect the onset of non-Markovian dynamics~\cite{lu_quantum_2010}, and plays a central role in quantum metrology. 
Quantum metrology involves distinguishing between quantum states $\rho_\param$ and $\rho_{\param+\delta\param}$ where $\delta\param$ is an infinitesimal change of the parameter $\param\in\mathbb R$ that characterizes the state ensemble generated by the signal of interest.
In quantum metrology, the QFI is related to the ultimate precision bound given by the Cram\'{e}r-Rao bound~\cite{braunstein_statistical_1994,braunstein_generalized_1996,cramer1999mathematical}. 
The bound is asymptotically saturated for optimal projective measurement, which also corresponds to the eigenstates of the {\it symmetric logarithmic derivative} (SLD)~\cite{braunstein_statistical_1994, holevo_probabilistic_2011} that defines the Bures metric. 

Although the role of QFI and SLD is crucial, their evaluation is notoriously difficult because of their nonlinear mathematical form. 
Among various formulas, a simplified version of the QFI for an adiabatic perturbation in a quantum many-body system can be derived~\cite{mihailescu_quantum_2024}. 
In general, calculating the QFI requires matrix inversions, which become significant numerical bottlenecks.
As a result, considerable effort has been devoted to developing methods for evaluating or bounding the QFI, paralleling advances in quantum metrology. 
The conventional approach, which relies heavily on eigendecomposition, faces a critical limitation: the Hilbert space dimension grows exponentially with the number of subsystems, making computations intractable for large, composite systems. 
To address these challenges, alternative formulations that avoid explicit diagonalization have been proposed~\cite{safranek_simple_2018,fiderer_general_2021}, although they still require certain matrix inversions. 
Other practical approaches provide QFI bounds using auxiliary measures such as the truncated QFI~\cite{sone_generalized_2021,beckey_variational_2022}, purity loss~\cite{modi_fragile_2016,yang_probe_2020}, and the Krylov shadow method~\cite{zhang_krylov_2025}.
Recent experimental advances have used noisy intermediate-scale quantum (NISQ) devices to estimate these bounds~\cite{sone_generalized_2021,beckey_variational_2022,dellanna_quantum_2025}, enabling scalable implementation on physical quantum platforms. 
In recent decades, {\it tensor network}~(TN)~\cite{bridgeman_hand-waving_2017,orus_tensor_2019} methods have been shown to provide efficient approximations for relevant classes of states~\cite{white_density_1992,haegeman_time-dependent_2011,orus_tensor_2019}, e.g., ground states of gapped Hamiltonians and thermal states of a local Hamiltonian~\cite{eisert_colloquium_2010,alhambra_locally_2021}. 
Among them, the one-dimensional ansatz of the {\it matrix product state} (MPS) (also known as the {\it matrix product operator} (MPO) for quantum operators) provides a compact representation of quantum states and operators. 

Although the literature on general-purpose TNs is extensive, there is considerable room for improvement regarding their application to quantum metrology. 
Among the existing approaches, Ref.~\cite{hauru_uhlmann_2018} introduced a TN-based procedure to calculate quantum fidelity between connected subsystems of partially traced pure-state matrix product states. 
For mixed states, a more general method was proposed in Ref.~\cite{chabuda_tensor-network_2020}, where both the quantum Fisher information (QFI) and the symmetric logarithmic derivative (SLD) are obtained through variational optimization. 
The original implementation of the approach was released as the open-source package TNQMetro~\cite{chabuda_tnqmetro_2022} (followed by QMetro++~\cite{dulian_qmetro_2025}). 
More recently, Ref.~\cite{yang2025quantumcramerraoprecisionlimit} presented an efficient framework for evaluating the quantum Cramér-Rao bound in continuously monitored systems subject to general Markovian or non-Markovian noise.
In this paper, we introduce a novel approach suitable for evaluating the QFI and SLD from their integral forms. 
Although integral formulations are not new, our approach leverages tensor networks that enable scalable computations for large-scale systems. 
Importantly, the procedure can be efficiently implemented using imaginary-time evolution and basic tensor network algebra. 
This makes our approach directly applicable utilizing open source packages that include general MPS implementation, such as YASTN~\cite{yastn_scipost,yastn_codebase}, TeNPy~\cite{hauschild2018efficient} or ITensor~\cite{fishman2022itensor}. 
In this paper, we analyze our method in the context of convergence time, the cost of numerical implementation, and numerical errors. 

This paper is organized as follows. 
In Sec.~\ref{sec:preliminaries} we introduce the concept of SLD and QFI and their integral forms that are fundamental for our method. 
Section~\ref{sec:metrology} provides an essential background on quantum metrology, which will be used to analyze our method. 
In Sec.~\ref{sec:implementation} we discuss the basic scheme for approximate the SLD and QFI integrals. 
Section~\ref{sec:convergence} analyzes the convergence properties of the method and derives analytical bounds for the associated error.
Section~\ref{sec:model} delves into technical details, including tensor network implementation, computational complexity, and error analysis. 
For illustration, we use our method to study quantum metrology where we employ a finite-temperature {\it transverse field Ising model} (TFIM) as a probe to estimate an external field.
The model provides an ideal benchmark example, as its established physics helps correlate the convergence of our method with the underlying physical properties of the probe. 

\section{Preliminaries}
\label{sec:preliminaries}

In quantum information, an infinitesimal distance between quantum states is defined by the Bures metric~\cite{bures_extension_1969,uhlmann_metric_1992,holevo_probabilistic_2011}
\[\label{eq:bures}
d_B^2(\rhoParam{\param}, \rhoParam{\param+\delta\param})  = \frac{1}{2}\tr{\left[\dT{\rhoParam{\param}}\sld\right]},
\]
where $\dT {\rhoParam{\param}}  \equiv \frac{\partial{\rhoParam{\param}}}{\partial\param}$ is a derivative of the density matrix along $\param$ which parameterizes the ensemble of quantum states $\rhoParam{\param}$. 
The {\it symmetric logarithmic derivative}~(SLD)~\cite{braunstein_statistical_1994, holevo_probabilistic_2011} is defined by the equation
\[\label{eq:sld_def}
\dT{\rhoParam{\param}} = \frac{1}{2}\left({\rhoParam{\param}} \sld + \sld {\rhoParam{\param}}\right) ,
\]
where $\sld$ is the SLD operator. 
The geometric distance defined by the Bures metric is proportional to the quantum Fisher information. 
The {\it quantum Fisher information} (QFI)~\cite{helstrom_quantum_1969,paris_quantum_2009,holevo_probabilistic_2011} is expressed in the following form
\[\label{eq:qfi}
\qfi=\tr{\left[\dT{\rhoParam{\param}}\sld\right]}
\]
or equivalently $\qfi=\tr{\left[{\rhoParam{\param}} \sld^2\right]}$. 
The cases where the rank of the density matrices changes with the parameter $\param$ and the relation between the Bures metric and the QFI is more intricate are discussed in Refs.~\cite{safranek_discontinuities_2017,rezakhani_continuity_2019,seveso_discontinuity_2020}. 
Here, we will limit ourselves to the cases where they are directly proportional to each other and the SLD in Eq.~\eqref{eq:sld_def} is well defined. 

We build the foundation of our method by identifying the definition in Eq.~\eqref{eq:sld_def} as a continuous Lyapunov equation~\cite{lancaster_explicit_1970,simoncini_computational_2016}. 
The analytical solution of the SLD in Eq.~\eqref{eq:sld_def} can be expressed in terms of the so-called {\it Lyapunov integral}~\cite{paris_quantum_2009,liu_quantum_2016}
\[\label{eq:sld_integral}
\sld = 2 \int_0^\infty d\arg \, e^{-{\rhoParam{\param}} \arg} \dT {\rhoParam{\param}} \, e^{-{\rhoParam{\param}} \arg} ,
\]
and the QFI reads
\[\label{eq:qfi_integral}
\qfi = 2 \int_0^\infty d\arg \, \tr[ \dT {\rhoParam{\param}} \, e^{-{\rhoParam{\param}} \arg} \dT {\rhoParam{\param}} \, e^{-{\rhoParam{\param}} \arg}] .
\]
The formal solution presented in Eq.~\eqref{eq:sld_integral} applies to the asymptotically stable Lyapunov equation and is equivalent to the pseudoinverse.
The integral form remains valid when the density matrix is not full-rank, although additional regularization is necessary~\cite{safranek_discontinuities_2017,rezakhani_continuity_2019,huang_novel_2021}. 
In particular, it can be shown that the QFI matrix of any state can be calculated as a limit of the QFI matrix of a full-rank state~\cite{safranek_discontinuities_2017}, e.g. $\qfi = \lim_{\nu\rightarrow 0} \, (1-\nu)\rho_\param + \nu\mathbbm{1}$ and $\nu\in(0,1)$. 
An extended discussion of the SLD can be found in App.~\ref{sec:convergence_general}.
In this work, we restrict our attention to the cases where the SLD is well defined by the definition in Eq.~\eqref{eq:sld_def} and the integral solution in Eq.~\eqref{eq:sld_integral} is valid.

A key advantage of the integral form is that it eliminates the need for an exact diagonalization of the density matrices. 
Both matrix multiplication and diagonalization scale cubically with the dimension of the density matrix, which itself grows exponentially with system size in the full Hilbert space, making computations for large systems infeasible. 
However, in contrast to exact diagonalization, the Lyapunov integral is particularly well-suited for numerical evaluation using tensor networks, where $e^{-\rhoParam{\param}\arg}$ can be efficiently approximated by Krylov methods or Trotter decomposition. 
In practice, such exponentials are applied using standard algorithms for time evolution, e.g. TDVP~\cite{haegeman_time-dependent_2011}, and all operators can be represented as matrix product operators. 
By enabling local updates of these matrix product operators, this approach overcomes the exponential scaling problem and paves the way for large-scale simulations. 
Furthermore, we note that the QFI formulation in Eq.~\eqref{eq:qfi_integral} circumvents the necessity for explicit construction of the SLD operator and can be treated with standard methods for numerical integration. 
Although our method is more broadly applicable, we will discuss it in the application to unitary quantum metrology. 

\section{Quantum metrology}
\label{sec:metrology}

\begin{figure}[t!]
    \centering
    \includegraphics[width=\columnwidth]{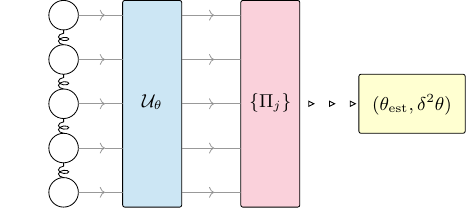}
    \caption{{\bf Typical procedure for quantum metrology.}
    The generally entangled input probe (white) is prepared in arbitrary quantum protocol.
    The probe is exposed to the signal generated by the unitary $U_\param$ that encodes the parameter $\param$ (blue). 
    After the encoding the probe is measured with the positive operator-valued measure $\{\Pi_j\}$ (red). 
    The measurement data are analyzed with statistical processing (yellow) to estimate the parameter value $\param_{\rm est}$ and its variance $\delta^2\param$. 
   }
    \label{fig:fig1}
\end{figure}

Quantum metrology aims to leverage entanglement and superposition in order to improve the precision of the estimation of physical parameters.
A typical estimation protocol follows a preparation-encoding-measurement sequence as visualized in Fig.~\ref{fig:fig1}: a generally entangled quantum probe is first prepared; then the probe is exposed to the signal generated by the unitary $U_\param$ that encodes the parameter $\param$; later the probe is measured with the positive operator-valued measure defined by the set $\{\Pi_j\}$; the measurement data are analyzed to calculate the estimated value of the parameter $\param_{\rm est}$ and the error of the value given by the variance $\delta^2\param$. 

The bound on the estimation precision, given large measurement statistics, is fixed by 
the celebrated {\it (quantum) Cram\'{e}r-Rao bound}~\cite{helstrom_minimum_1967,braunstein_statistical_1994,braunstein_generalized_1996,cramer1999mathematical,paris_quantum_2009}
\begin{equation}\label{eq:cramerrao}
    \delta^2\param \geq \frac{1}{\qfi} ,
\end{equation}
where $\delta^2\param$ is the variance of the estimated parameter $\param$ and $\qfi$ is the {\it Fisher information} (FI) calculated for the probability density function of measurement outcomes~\cite{cramer1999mathematical}. 
The {\it quantum Fisher information} (QFI) is defined as the maximum value of the Fisher information achievable across all possible measurement strategies, and it is also closely related to the Bures metric, as discussed in Sec.~\ref{sec:preliminaries}. 
The highest estimation precision is achieved through the optimal measurement, which is determined by the SLD operator defined in Eq.~\eqref{eq:sld_def}~\cite{helstrom_minimum_1967,braunstein_statistical_1994,braunstein_generalized_1996,paris_quantum_2009}. 
Specifically, optimal projective measurement corresponds to the eigenvectors of the SLD operator. 

\section{Integration method}
\label{sec:implementation}

The expressions for QFI and SLD in Eq.~\eqref{eq:sld_integral} and Eq.~\eqref{eq:qfi_integral} respectively are approximated by truncated integrals of the form
\[\label{eq:BT0}
\sld(\argLim) = 2 \int_0^\argLim d\arg \, \B(\arg) ,
\]
where $\argLim$ is the upper limit of integration and the SLD integrand reads
\[\label{eq:Bs}
\B(\arg) = e^{-{\rhoParam{\param}} \arg} \dT {\rhoParam{\param}} \, e^{-{\rhoParam{\param}} \arg} .
\]
Similarly, for the QFI in Eq.~\eqref{eq:qfi_integral} we write the truncated integration
\[\label{eq:qfi_integral_trunc}
\qfi(\argLim) = 2 \int_0^\argLim d\arg \, \dqfi(\arg)
\]
where the QFI integrand is given by
\[\label{eq:qfi_integrand}
\dqfi(\arg) = \tr[\dT{\rhoParam{\param}}\B(\arg)] = ||\B(\arg/2)||_2^2 ,  
\]
and $||\B(\arg/2)||_2^2 = \tr[\B(\arg/2)\B(\arg/2)]$ is the square of a 2-norm of the operator, i.e. $||{\hat O}||_2 = \sqrt{\tr[{\hat O}{\hat O}^\dagger]}$ ($\B(\arg/2)$ is a Hermitian operator). 
Notice that using $\dqfi(\arg) = ||\B(\arg/2)||_2^2$ takes operators evaluated at $\arg/2$, which reduces the computational cost used for propagation of the integrand. 
The integral truncation is warranted, as the integral converges with increasing $\arg$, see also App.~\ref{sec:convergence_time} for further details.
In Sec.~\ref{sec:convergence}, we discuss the bounds on the convergence error. 

To numerically implement this, it is also necessary to convert the integral into a summation. This discretization is established using finite integration steps $\Darg_j$, where these steps satisfy condition $\sum_j\Darg_j=\argLim$. 
The sum rule can be chosen in a similar way as in standard numerical integration schemes, e.g., according to the quadrature rule, trapezoid rule, or higher-order Runge-Kutta (RK) methods~\cite{ZHENG2017361}. 
The integrand is propagated in the argument $\arg$ by iterative updates of the integrand operator in Eq.~\eqref{eq:Bs}, such that $\B(\arg+\Darg) = e^{-{\rhoParam{\param}} \Darg} \B(\arg) e^{-{\rhoParam{\param}} \Darg}$. 
This step can be easily implemented using standard TNs' methods for time evolution, such as TDVP~\cite{haegeman_time-dependent_2011,haegeman_unifying_2016} or Krylov methods~\cite{zaletel_time-evolving_2015}. 
Having obtained the operator $\B(\arg)$, it is accumulated according to the sum rule for the SLD operator in Eq.~\eqref{eq:BT0}. 
From the SLD we can calculate the QFI using the expression in Eq.~\eqref{eq:qfi}. 
Importantly, the QFI can be directly integrated using the expression in Eq.~\eqref{eq:qfi_integral_trunc} without the necessity of explicitly constructing the SLD operator. 
In this context, the QFI integrand can be aggregated in a manner analogous to the SLD or integrated by fitting during the post-processing phase.

\section{Convergence}
\label{sec:convergence}

The Lyapunov integral spreads the derivative of the probe state across its spectral decomposition, illustrating how parameter dependence permeates the entire density matrix. 
This formulation is well suited for not only numerical calculations, but also offers a clear physical insight into the sensitivity of the probe to changes in parameters. 
In particular, the QFI integrand reads
\[\label{eq:B_spectral}
\B(\arg) = \sum_{ij}{|\dT{\rhoParam{\param}}_{ij}|^2} e^{-(\lambda_i+\lambda_j)\arg} ,
\]
where $\rhoParam{\param}=\sum_j\lambda_j\ket{j}\bra{j}$ and $\dT{\rhoParam{\param}}_{ij}=\bra{i}\dT{\rhoParam{\param}}\ket{j}$. 
Each term in the sum exhibits a unique decay rate determined by the corresponding pair of eigenvalues, $\lambda_i$ and $\lambda_j$. 
Furthermore, the amplitude of each term is governed by the state derivative $\dT{\rhoParam{\param}}_{ij}$; thus, if the encoding more strongly couples these terms, their contributions become correspondingly greater.
Exponential decay causes the largest eigenvalue contributions to fade away first, thereby exposing the next set of eigenvalues. 
As a result, the integrand provides information on how the derivative is distributed across the state density matrix. 

Since the QFI integrand is always positive, the truncation of the integrand gives the lower bound on the exact QFI and the truncated solution satisfies
\[
\qfi(\argLim_1)\leq\qfi(\argLim_2)\leq\qfi(\infty) ,
\]
where $\argLim_1\leq \argLim_2\leq\infty$. 
The convergence rate of the QFI integral is influenced by both the cutoff parameter $\argLim$ and the spectral properties of the density matrix. 
As $\argLim$ increases, terms that involve smaller eigenvalues become more significant, leading to a slower decay of the integrand in Eq.~\eqref{eq:B_spectral}. 
Consequently, contributions with small amplitudes require long integration time. 
However, we note that the QFI integrand in Eq.~\eqref{eq:B_spectral} is proportional to the square of the eigenvalues (this is, up to the velocities appearing in the full expression $\bra{i}\partial_\param{\rhoParam{\param}}\ket{j} = (\partial_\param\lambda_j) \delta_{ij} + \lambda_j \bra{i}\partial_\param j\rangle + \lambda_i \bra{\partial_\param i}j\rangle$) and, therefore, the total integral over these terms scales as the amplitude of the eigenvalues themselves. 
Intuitively, if small eigenvalues are a minority of the eigenspectrum, the QFI is well approximated by a moderate $\argLim$. \\

A general bound on the convergence error is difficult to calculate. 
In this section, we focus on the scenario for typical for quantum metrology as explained in Sec.~\ref{sec:metrology}. 
While the protocol for quantum metrology can be generalized, we will focus on the unitary encoding where the probe undergoes the following transformation
\[\label{eq:encoding}
{\rhoParam{\param}} = e^{-\im \param A} \rhoParam{0} \, e^{\im \param A} ,
\]
where $\rhoParam{0}$ is the state of the prepared probe, $\param$ is the encoded parameter (often the parameter will be the accumulated phase, i.e. $\param=\Theta t$ where $\Theta$ is the encoded parameter and $t$ is an interrogation time), and $A$ is the encoded operator generating the encoding unitary $U_\param = e^{-\im \param A}$. 
There, the state derivative over the parameter $\param$ reads
\[
\dT{\rhoParam{\param}}=-\im[A,\rhoParam{\param}]. 
\]
A relevant measure of convergence is the relative error of the QFI, which is defined as
\[\label{eq:epsilon}
\relErr(\argLim) = \frac{\qfi(\infty) - \qfi(\argLim)}{\qfi(\infty)} .
\]
where, in its spectral representation, the convergence error is expressed as
\[
\label{eq:errqfi_spectral}
\qfi(\infty) - \qfi(\argLim) = 2 \sum_{ij}\frac{|A_{ij}|^2 (\lambda_i-\lambda_j)^2}{\lambda_i+\lambda_j} e^{-(\lambda_i+\lambda_j) \argLim} ,
\]
where $A_{ij}=\bra{i}A\ket{j}$ is an element of the encoding operator in the eigenbasis of density matrix. 
Regardless of the spectral properties of the density matrix, the error is bounded by the relation
\[\label{eq:worstbound}
\relErr(\argLim) \leq 2 \frac{\sum_{i,j\neq i} |A_{ij}|^2 }{\qfi(\infty)} {\min}[({e\argLim})^{-1}, e^{-\lambda_{\rm min}\argLim}],
\]
suggesting the convergence rate decays inversely with the integration cutoff $\argLim$. 
See App.~\ref{sec:convergence_time} for details of the derivation. 
The bound in Eq.~\eqref{eq:worstbound} is the worst-case scenario derived by assuming the maximum possible error contribution for each pair of $\lambda_i$ and $\lambda_j$ eigenvalues. 

Sacrificing generality, we can derive a bound for low-entropy states, such as low-temperature thermal states, where the state is well approximated by the ground state and the first excited state. 
Here, we demonstrate that the convergence error increases linearly with the entropy of the density matrix
\[\label{eq:boundLotTemp}
\relErr(\argLim) \lesssim e^{-\argLim/m} + \frac{c}{2} \frac{m}{n} \left(S(\rhoParam{\param})\frac{m}{2\log2}-m+1\right) ,
\]
where $S(\rhoParam{\param})$ is the von Neumann entropy, $c=\sum_{j\neq{GS, ES}}|A_{ES, j}|^2/\sum_{j\neq{GS}}|A_{GS, j}|^2$ is an encoding-dependent constant, $GS$ stands for the ground states with degeneracy $m$ and $ES$ stands for the excited states with degeneracy $n$. 
The theoretical bound in Eq.~\eqref{eq:boundLotTemp} implies that the convergence cutoff $\argLim$ should increase with the entropy making the numerical study more challenging. 
Likewise, the convergence time should increase with degeneracy. 
The complete derivation is provided in App.~\ref{sec:convergence_time_lowT}. 

The convergence of any state can be estimated by analyzing the principal components of the density matrix. 
These can be found numerically using standard variational algorithms for MPS such as DMRG~\cite{white_density_1992,orus_tensor_2019} or Krylov methods. 
After learning the first few dominant eigenvalues of the spectrum, we can explicitly write the QFI error in Eq.~\eqref{eq:errqfi_spectral} for the known set $\Lambda$ where $\lambda_i\in\Lambda$ are known. 
Typically, only the first few eigenvalues are found, making the corrections arising from unknown eigenvalues significant.
The correction coming purely from the remaining part of the spectrum is bounded by
\[\label{eq:dqfi_complement}
\relErr(\argLim)_{\tilde\Lambda}\leq \frac{4 c_{\tilde\Lambda}}{\qfi(\infty)} {p_{\tilde\Lambda}}(n_{\tilde\Lambda}-1) e^{-{2p_{\tilde\Lambda}}X/n_{\tilde\Lambda}} ,
\]
where we assumed that $X\leq \frac{1}{2\Lambda_{\rm min}}$, $\tilde\Lambda$ represents the set of unknown eigenvalues, $p_{\tilde\Lambda}=1-\tr[\Lambda]$ denotes the population of the unknown set $\tilde\Lambda$, $n_{\tilde\Lambda}=\dim[\rhoParam{\theta}] - \dim[\Lambda]$ is their portion within the Hilbert space, and $c_{\tilde\Lambda}=\max_{i,j\in\tilde\Lambda} |A_{ij}|^2$ is a constant related to the encoding.
For the full derivation, see App.~\ref{sec:convergence_spectrum}. 
Equation~\eqref{eq:dqfi_complement} demonstrates that the correction becomes more pronounced as $p_{\tilde\Lambda}$ and $n_{\tilde\Lambda}$ increase.
This suggests again that states of small entropy will be most suitable to efficiently compute the QFI using numerical integration. 

Notice that the truncation error can be reduced by extrapolating the QFI integrand. 
In this approach, we extract the decay rate in the window of $\arg\in[\argLim-D[\Darg], \argLim]$ at the end of the simulation assuming an exponential decay (or a sum of exponential decays). 
We extrapolate this decay from $\argLim$ up to $\infty$ and analytically integrate the fit by adding the residual integral to $\qfi(\argLim)$. 
The fitting procedure naturally disregards contributions that arise for $X \gg \argLim$, that is, exponentials associated with eigenvalues smaller than the fitted rate. 
This approximation remains valid because the overall impact of these terms is proportional to the amplitudes of the corresponding eigenvalues. 

\section{Results}
\label{sec:model}

Building on the information presented in the previous sections, we now develop and analyze our approach. 
We explore its properties, focusing on convergence and numerical implementation using tensor networks. 
To demonstrate the practical utility of our method, we apply it to a specific model in quantum metrology in Fig.~\ref{fig:fig1}, although the majority of our findings can be generalized. 

\subsection{Model}
\label{sec:convergenceTFIM}

We consider a paradigmatic example of quantum metrology in which a quantum probe is used to estimate the magnetic field modeled by the Hamiltonian
\[\label{eq:A}
A = \sum_{j=1}^N \z{j} ,
\]
where $\sigma^\alpha_{j}$ are standard $\alpha\in\{x,y,z\}$ Pauli operators for a spin $j$. 
The operator $A$ generates an encoding according to the scheme in Eq.~\eqref{eq:encoding} with the parameter value $\param=J$ where $J$ is treated here as the frequency unit of the model. 
To demonstrate a nontrivial application of our method, we examine estimation protocols using a mixed-state quantum probe. 
We consider an ensemble of thermal states of the {\it transverse field Ising model} (TFIM) with open boundary conditions described by the Hamiltonian
\[\label{eq:model}
H_\tf = - J \sum_{j=1}^{N-1} \x{j} \x{j+1} - \tf \sum_{j=1}^{N} \z{j} ,
\]
where $N$ is the number of spins, $J$ is the strength of the exchange interaction, and $\tf$ is the amplitude of the transverse field. 
The Hamiltonian generates an ensemble of thermal states parameterized by the transverse-field amplitude
\[\label{eq:thermal_probe}
\rhoParam{0} = \frac{e^{-\beta H_\tf}}{Z_{\tf, \beta}} ,
\]
where $[A,H_\tf]\neq 0$, $\beta$ is the inverse temperature and $Z_{\tf,\beta} = \tr{[e^{-\beta H_\tf}]}$ is the partition function. 
The probe ansatz allows to demonstrate the application of our approach to density matrices of various spectral properties. 
The TFIM has a rich phase diagram, with phases of degenerate and non-degenerate spectra, as well as the critical phase at $\tf_c=J$. 
As a well-studied model, the TFIM aids in the interpretation of results and physical intuition. 
The inverse temperature $\beta$ controls the purity of the thermal state and changes the distribution of the eigenvalues of the density matrix. 
Overall, the example covers a broad range of typical cases through a simple ansatz.

\begin{figure}[b!]
    \centering
    \includegraphics[width=\columnwidth]{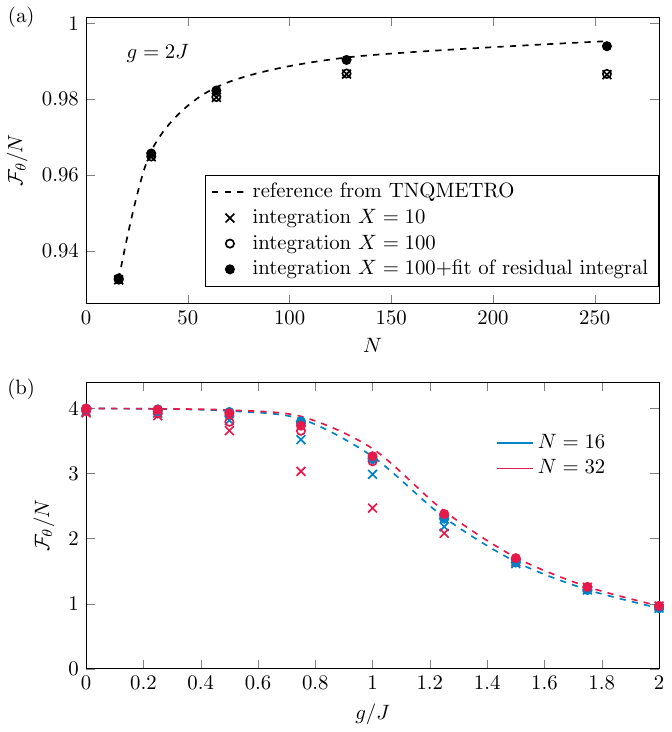}
    \caption{
    {\bf Phase estimation with thermal states of the TFIM.}
    Benchmark of the QFI obtained from truncated integration up to $\argLim$ (crosses and circles) compared to the reference value (dashed). 
    For $\argLim=100$ (dots) includes additional term from the fit of the residual integral obtained from the final $D[\Darg]=10$ range. 
    (a) Convergence in terms of system size at $\tf=2J$ for various system size and truncation cutoff $\argLim$. 
    (b) The benchmark of the QFI across the phase transition. 
    The convergence time increases around the critical point $\tf=J$ where the gap to the excited state becomes small. 
    Data obtained for encoding parameter $\param=J$, inverse temperature $\beta=4J^{-1}$ and bond dimension $D_{\rhoParam{\param}}\leq32$ for $N=16$ and $D_{\rhoParam{\param}}\leq64$ otherwise. 
    The integration obtained from the interpolation fit to the QFI integrand that was evaluated numerically with tensor networks. 
    The QFI integrand is calculated with the MPO ansatz of the bond dimension $D_\B\leq64$. 
    Reference data obtained with TNQMETRO~\cite{chabuda_tnqmetro_2022} using $D_\sld=16$. 
    }
    \label{fig:fig2}
\end{figure}

Figure~\ref{fig:fig2} shows the benchmark results of the QFI throughout the TFIM phase diagram. 
These results can be interpreted in the context of the TFIM Hamiltonian, where the spectral properties are controlled by the parameter $\tf$. 
At low temperature, convergence can be understood through the gap between the ground state and the first excited state. 
Here, the bound in Eq.~\eqref{eq:boundLotTemp} is the most relevant. 
The leading eigenvalue of the density matrix is the population in the ground state
\[\label{eq:GS_thermal}
\lambda_{\rm GS} = \frac{1}{m(1+e^{-\beta\Delta})} , 
\]
where $m$ is a degeneracy of $GS$ and $\Delta$ is the energy gap between $GS$ and $ES$. 
The amplitude in Eq.~\eqref{eq:GS_thermal} decreases with a reduction of both $\beta$ and $\Delta$. 
At the same time, the population of the excited state increases with the changes in these parameters. 
A higher degeneracy of both states leads to a lower population per individual eigenstate.
In the disordered phase ($\tf > J$), the ground state of the TFIM is non-degenerate. 
The energy gap to the excited state decreases as the system approaches the critical point at $\tf_c = J$, approximately linearly $\Delta = 2J(\tf - 1)$ for large systems. 
In contrast, in the ordered phase ($\tf < J$), the ground state is two-fold degenerate. 
The gap scales similarly to the disordered phase, with $\Delta = 2J(1 - \tf)$ for large systems reducing to zero at the critical point. 
Finally, the energy gaps decrease with increasing system size $N$, resulting in slower convergence for larger systems. 

\begin{figure}[b!]
    \centering
    \includegraphics[width=\columnwidth]{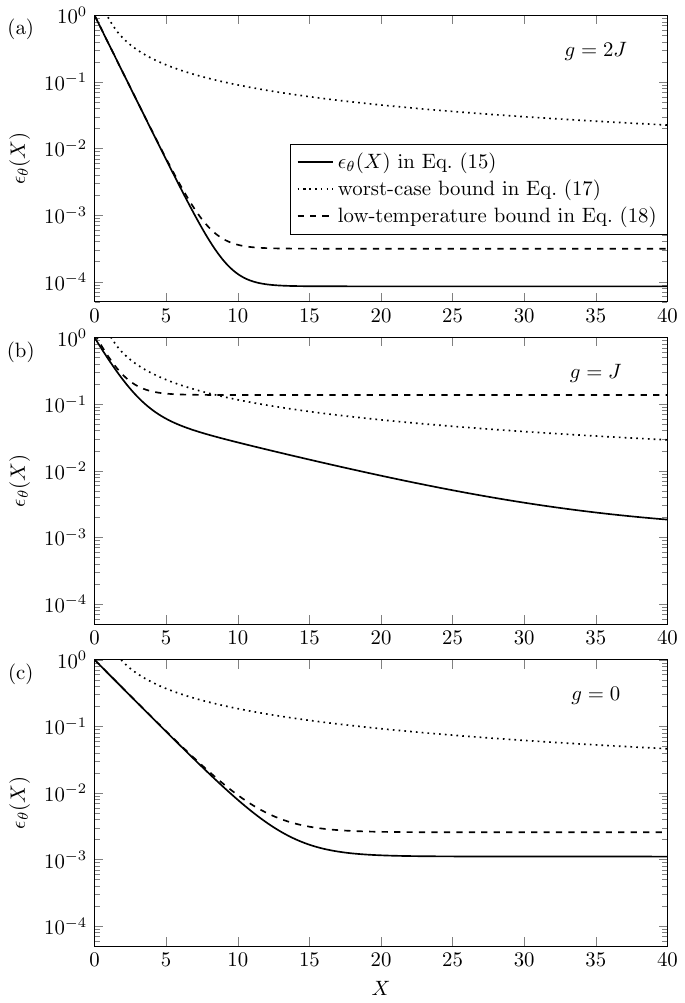}
    \caption{
    {\bf Convergence of the QFI integral.}
    Convergence of our method for the TFIM across the phase diagram with an example for the disordered phase $\tf=2J$ in (a), the critical point $\tf=J$ (b) and the ordered phase $\tf=0$ in (c). 
    Relative error in Eq.~\eqref{eq:qfi_integral} (solid lines) is plotted together with the low-temperature bound in Eq.~\eqref{eq:boundLotTemp} (dashed) and the worst-case bound in Eq.~\eqref{eq:worstbound} (dotted). 
    Data obtained for the thermal state of the TFIM in Eq.~\eqref{eq:model} with $N=6$ for $\beta=4J^{-1}$ by exact integration of the QFI in Eq.~\eqref{eq:qfi_integral}. 
    }
    \label{fig:fig3}
\end{figure}

In Fig.~\ref{fig:fig2}a, we show the results for $\tf=2J$. 
There, the Hamiltonian exhibits a macroscopic energy gap, and the ground state dominates the integral. 
Integrating up to $\argLim=10$ already captures most of the QFI, although a notable residual remains. 
The magnitude of the residual is proportional to the population of the first excited state and increases with system size as the energy gap narrows. 
The convergence rate of the residual integrand is proportional to the population of the excited level, which is small.
This results in a small improvement between $\argLim=10$ and $\argLim=100$, however, the residual is well approximated by a single exponent and can be easily integrated by fit.
The QFI with the fit approximates well the reference value. 
Figure~\ref{fig:fig2}b shows the benchmark across the phase transition. 
We observe that the convergence time increases around the critical point. 
This can be easily interpreted as a consequence of the decreasing gap. 
Within both the disordered and ordered phases, the convergence time remains moderate. 
However, the ordered phase displays a marginally greater error, which is due to the degeneracy of its ground state.

In Fig.~\ref{fig:fig3}, we present an explicit relation of the convergence error in Eq.~\eqref{eq:epsilon} as a function of the cutoff $\argLim$. 
The plot shows results for a small system where the bounds can be calculated explicitly. 
These plots clearly illustrate how the spectral properties influence the convergence rate. 
In particular, it highlights how the convergence of the dominant QFI contribution depends on the ground-state degeneracy and energy gap, while the residual tail is associated with contributions from excited states. 
Figure~\ref{fig:fig3} shows the convergence of the relative error compared to the bounds in Eq.~\eqref{eq:worstbound} and Eq.~\eqref{eq:boundLotTemp}. 
The general bound in Eq.~\eqref{eq:worstbound} provides a worst-case scenario, which is visually less tight. 
The low-temperature bound in Eq.~\eqref{eq:boundLotTemp} predicts an exponential convergence of the leading contribution with rate $e^{-\argLim/m}$, where $m$ denotes the ground-state degeneracy. 
This bound agrees well with the exact error in the dominant contribution, but does not capture the residual decay which is replaced by the entropy-dependent constant. 

\subsection{Tensor network implementation}

Tensor networks provide a powerful framework for encoding quantum states and operators. 
For one-dimensional tensors and open boundary conditions of the model in Eq.~\eqref{eq:model} we choose an ansatz of {\it matrix product operator} (MPO)~\cite{verstraete_matrix_2004,orus_tensor_2019}. 
In this representation, operators are decomposed into a series of rank-$4$ tensors, where two indices correspond to local {\it bra} and {\it ket} dimensions, while the others are so-called bond dimensions that encode correlations between local variables. 
The tensors form a one-dimensional network with a trivial dimension for the terminal bond dimension. 
The compression of the MPO relies on the Schmidt decomposition in subsequent bipartitions. 
The decomposition splits the operator into the sum $K = \sum_j \kappa_j L_j \otimes R_j$, where $\kappa_j$ are the Schmidt values and $L_j$ and $R_j$ are associated Schmidt left and right operators, respectively, such that $\tr[L_iL^\dagger_j]=\tr[R_iR^\dagger_j]=\delta_{ij}$. 
To reduce the bond dimension, we truncate the Schmidt decomposition at the Schmidt value cutoff $\SVtol$ such that $\kappa_j\geq\SVtol$. 
Alternatively, we keep the maximal maximal bond dimension $D_K$ which fixes a number of significant contributions that we keep. 

In this paper, the thermal state is obtained using imaginary time evolution with the Hamiltonian starting from the maximally mixed, i.e. $\beta=0$, state. 
The bond dimension exhibits a significant increase near the critical point, where the state is highly correlated. 
In terms of temperature, the bond dimension of the MPO representation scales proportionally to the inverse temperature~\cite{korepin_universality_2004,alhambra_locally_2021, kuwahara_improved_2021}. 
For TFIM, the operator space entanglement entropy (OSEE) was reported to increase logarithmically in the inverse temperature $\beta$~\cite{znidaric_complexity_2008}. 
The OSEE is not directly related to the bond dimension for a general model but gives quantitative measure of the amount of correlations between subsystems and predicts increasing hardship of the representation of the critical state. 
For non-critical states, the computational scaling is much more favorable, and the OSEE quickly saturates with $\beta$. 

\subsection{Unitary encoding}
\label{sec:unitary}

For unitary encoding scenarios, we can take advantage of the structural properties of the state derivative to formulate an alternative integration methodology.
In this case, the integration can be moved to the encoding operator, mirroring the time evolution process in the Heisenberg picture~\cite{hartmann_density_2009,clark_exact_2010,cilluffo2024simulatinggaussianbosonsampling}. 
This technique offers a different perspective, potentially simplifying computations for certain types of problem. 

When substituting the state derivative $\dT{\rhoParam{\param}}=-\im[A,{\rhoParam{\param}}]$ into the integrand in Eq.~\eqref{eq:Bs}, we commute the exponents and ${\rhoParam{\param}}$ and move the propagation from the state derivative to the encoding operator $A$. 
The SLD integrand becomes
\[\label{eq:B4A}
\B(\arg) = -\im[\A(\arg), {\rhoParam{\param}}] , 
\]
where
\[\label{eq:As}
\A(\arg) = e^{-{\rhoParam{\param}} \arg} A\,e^{-{\rhoParam{\param}} \arg}
\]
is the integrand for the encoding operator. 
The integration gives the SLD operator
\[\label{eq:sld4A}
\sld(\argLim) = -\im[\K(\argLim), {\rhoParam{\param}}] ,
\]
where
\[\label{eq:AT0}
\K(\argLim) = \int_0^\argLim d\arg \, \A(\arg)
\]
is the integrated encoding operator. 

\begin{figure}[t!]
    \centering
    \includegraphics[width=\columnwidth]{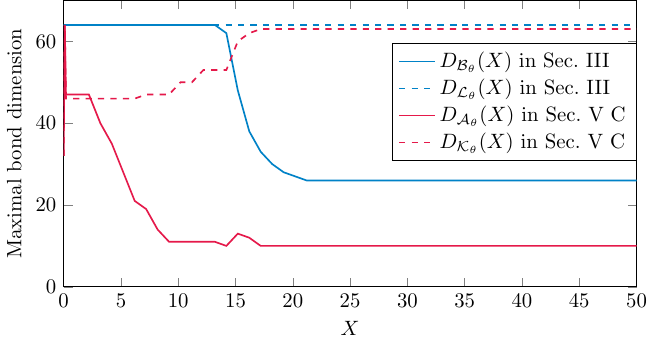}
    \caption{{\bf Variations for unitary encoding. }
    Comparison of the bond dimension obtained under the cutoff on Schmidt values $\SVtol=10^{-9}$ and maximal bond dimension $D\leq64$. 
    The figure compares the standard QFI integration introduced in Sec.~\ref{sec:implementation} (blue) and integration of the encoding in Sec.~\ref{sec:unitary} (red). 
    The integrand operators $\B(\argLim)$ in Eq.~\ref{eq:Bs} and $\A(\argLim)$ in Eq.~\ref{eq:As} (solid lines) are compared to the integrated operators $\sld(\argLim)$ in Eq.~\eqref{eq:BT0} and $\K(\argLim)$ in Eq.~\eqref{eq:AT0} (dashed). 
    Data obtained for $N=16$, $\tf=2J$, $\beta=4J^{-1}$, bond dimensions $D_{\rhoParam{\param}}\leq32$ truncation cutoff $\SVtol=10^{-9}$ and integration step $\Darg=0.1$.
    }
    \label{fig:fig4}
\end{figure}

Figure~\ref{fig:fig4} shows the bond dimension for each of the approaches. 
Working directly with the operator can benefit from a lower bond dimension, especially if the operator is local and has a low bond dimension. 
However, to obtain the QFI and SLD we need to perform a contraction between multiple MPOs. 
Complex contractions contribute to computational cost and can become limitations if the dimension of the bond is not controlled. 
Therefore, operator-based integration is advised for the SLD evaluation and discouraged for the direct QFI integration, where many contractions need to be computed. 
An extended comment on both methods for unitary encoding can be found in App.~\ref{sec:operator_vs_general}. 

\subsection{Computational cost}
\label{sec:TNcost}

In this paper, we compare the implementation costs of two methods: our method with integration, where the dominant cost arises from time evolution to propagate the integrand in Eq.~\eqref{eq:Bs}, and the variational approach, where the primary cost comes from computing the pseudoinverse to solve the linear problem for the local SLD update. 
In this paper, for time evolution, we use {\it time-dependent variational principle} (TDVP)~\cite{haegeman_time-dependent_2011,haegeman_unifying_2016} that allows us to evolve the MPO under an arbitrary generator. 
This is in contrast to methods such as TEBD~\cite{vidal_efficient_2003,vidal_efficient_2004} which is preferred for local generators. 
An extended discussion on the cost for the integration and variational approach can be found in App.\ref{sec:computational_cost} and App.~\ref{sec:varia}, respectively. 
In both approaches, we encode all operators using the ansatz of {\it matrix product operator} (MPO)~\cite{verstraete_matrix_2004,orus_tensor_2019} with the open-boundary condition. 

\begin{figure}[t!]
    \centering
    \includegraphics[width=\columnwidth]{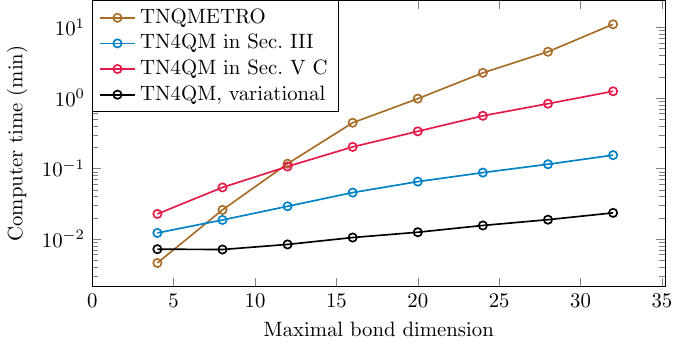}
    \caption{{\bf Cost of single update.}
    Comparison of the computational time for a single update obtained for the integration approach with the integrand $\B$, see Sec.~\ref{sec:implementation} (blue), $\A$, see Sec.~\ref{sec:unitary} (red), the variational approaches in the original implementation TNQMETRO~\cite{chabuda_tnqmetro_2022} (brown), and based on the Krylov method, see App.~\ref{sec:varia} (black). 
    The computational time is the smallest for our variational method implemented in TN4QM where Krylov methods are used reduce dimension of the problem we solve for local update. 
    Then, there is the $\B(\arg)$-based integration that takes simple overlaps in Eq.~\eqref{eq:qfi_integrand} to evaluate the integrand. 
    After that, the $\A(\arg)$-based integration is mode costly due to higher order overlaps in Eq.~\eqref{eq:B4A}. 
    Finally, the TNQMETRO is generally the most expensive because the preudoinverse is applied directly on effective operators that has a large dimension. 
    Data obtained using TNQMETRO~\cite{chabuda_tnqmetro_2022} and TN4QM~\cite{tn4qm} for $N=16$, $\tf=2J$, $\beta=4J^{-1}$, bond dimensions $D_{\rhoParam{\param}}\leq32$ and integrand $D_\B\leq64$, integration step $\Darg=0.1$. 
    Computer time obtained on a single CPU core. 
    }
    \label{fig:fig5}
\end{figure}

The numerical cost of a single update for both methods increases squared with the bond dimension of the density matrix, that is, $\mathcal{O}(D_{\rhoParam{\param}}^2)$ where $D_{\rhoParam{\param}}$ is a bond dimension of ${\rhoParam{\param}}$. 
The cost increases as $\mathcal{O}(D_\sld^3)$ for the variational method, which relies on the SLD operator $\sld$, and scales as $\mathcal{O}(D_{\B/\A}^3)$ for the integration approach, which relies on the integrand $\B/\A$. 
We remind the reader that the cost of the variational approach in the original implementation in Ref.~\cite{chabuda_tnqmetro_2022} is further increased to $\mathcal{O}(D_\sld^6)$ scaling due to applying pseudoinverse on the effective operators directly. 
In this paper, we suggest an alternative approach that relies on the Krylov methods. 
In our approach we solve the linear problem for the variational update in the auxiliary basis of Krylov vectors, see also App.~\ref{sec:varia}.  
This mitigates the extreme scaling for the pseudoinverse, returning the computational scaling to $\mathcal{O}(D_\sld^3)$ in terms of SLD. 
Figure~\ref{fig:fig5} presents a comparison of the computational cost for a single update in all the cases we discussed. 

Importantly, the total computational cost of each approach is determined by multiplying the cost of a single update by the number of updates needed to achieve convergence.
We numerically observe that the variational approach usually converges after a few or a few dozens of sweeps, while the integration usually uses more sweeps. 
For the integration approach, the number of steps is related to the convergence time $\argLim$ that was discussed in Sec.~\ref{sec:convergence}. 
There, we noticed that in the low temperature limit the convergence time increases with the state entropy. 
This relation does not apply for the variational approach. 
The number of steps to reach $\argLim$ using a fixed integration step $\Darg$ leads to the $\mathcal{O}(\argLim/\Darg)$ scaling, however, this can be changed when applying the adaptive approach in App.~\ref{sec:adaptive}. 
The truncation of the integrands $\B/\A$ can change with the integral argument. 
Although the exact behavior of the integrands is difficult to predict, computational efficiency can always be enhanced through weighted compression. 
In this approach, we apply stronger truncation in regions where the amplitude of the integrand decreases substantially. 
By adopting more aggressive truncation, where appropriate, we concentrate computational resources on the most significant contributions of the integrand. For further details, see App.~\ref{sec:computational_cost}.

Finally, we note that, much like the comparison of methods to obtain the ground state through variational techniques and through imaginary-time evolution~\cite{white_density_1992,haegeman_unifying_2016}, directly comparing integration and variational methods proves to be challenging because their underlying principles differ fundamentally. 
In general, we suggest to decide on the method based on the properties of a given problem. 
For example, when the state is highly mixed, the integration of the QFI will converge very slowly, while the variational methods will reach the solution directly. 
On the other hand, when the TNQMETRO is compared with our method through integration, the benefit of the latter is clearly manifested in the computational cost of individual updates even when the convergence time for the integration is significant. 
Finally, we remind the reader that, contrary to the variational methods, the QFI integration avoids direct evaluation of the SLD operator. 
Therefore, if the bond dimension of SLD is large, the integration method may benefit from lower computational cost, however, the bond dimension of the integrand is hard to determine. 

\subsection{Numerical errors}
\label{sec:errors}

The numerical error of the approach splits into two sources: the error of numerical integration and the error of tensor network representation. 
The error of numerical integration is related to integral truncation as discussed in Sec.~\ref{sec:convergence}. 
Here, we discuss the remaining sources of error. 

One of the sources of error is the accuracy with which the integrand is evaluated. 
In this paper, the integrand is propagated using the $2$nd-order TDVP algorithm for which the error scales as $\mathcal{O}(\Darg^2)$, where $\Darg$ is a discrete step for propagation of the integrand. 
To improve the accuracy of this step, we can apply higher-order TDVP procedures, however, this comes at increased computational cost. 
The second type of error is an accuracy of the numerical integration which depends on the integral discretization. 
For this step, we can employ standard numerical integration schemes, such as a quadrature rule, trapezoid rule, or higher-order Runge-Kutta (RK) methods~\cite{ZHENG2017361}. 
The error of the RK-$k$ approach scales as $\mathcal{O}(\Darg^k)$ where $\Darg$ is a discrete integration step. 
For QFI-focused calculations, the integral can be evaluated in post-processing by taking the integrand as the input. 
This allows for flexible application of integration methods using fitting or interpolation procedures. 
In contrast to QFI, the SLD operator cannot be fitted externally and the integration has to be applied using the sum rule. 
The integration by a sum can be optimized by adopting an adaptive approach, see App.~\ref{sec:adaptive}, where the integration step is dynamically adjusted based on variations of the QFI integrand. 
The approach allows to control the integration error under the fixed tolerance at the cost of monitoring the QFI integrand. 
The adaptive procedure naturally allows step sizes to grow with increasing $\arg$ as the integrand becomes flatter and flatter. 
However, directly coupling the step size for integration to the step size for integrand propagation imposes a practical limitation, since the latter must be sufficiently small for TDVP. 
This restriction can be easily lifted when the step sizes are decoupled, but it may not be a significant gain since the TDVP cost remains dominant. 

\section{Conclusions}

In this paper, we introduce a numerical method for evaluating the quantum Fisher information (QFI) and the symmetric logarithmic derivative (SLD) using the Lyapunov integral approach. 
The convergence of this integral is governed by the spectral properties of the density matrix, typically leading to an increased computational cost for states with higher entropy. 
To address scalability, we propose a tensor network-based implementation that leverages only time-evolution algorithms and basic tensor algebra, making it compatible with most existing tensor network toolboxes. 
We analyze the convergence and computational costs of our method in the context of quantum metrology. 
Benchmarking demonstrates that our approach enables QFI calculations for system sizes that are intractable with standard exact diagonalization techniques. 

Efficient QFI evaluation for complex states not only deepens our understanding of quantum resources, but also supports practical applications in quantum sensing and estimation.
Our framework can be naturally extended to noisy metrological scenarios, where open system dynamics replace unitary evolution~\cite{huelga_improvement_1997,escher_general_2011,demkowicz-dobrzanski_elusive_2012,smirne_ultimate_2016,altherr_quantum_2021}. 
Although this work focuses on quantum metrology, the applicability of QFI is broader, encompassing studies of non-Markovianity~\cite{lu_quantum_2010}, quantum criticality~\cite{zanardi_quantum_2008}, and even contexts such as quantum gravity~\cite{lashkari_canonical_2016}. 
More generally, our approach can serve as a numerical solver for the continuous Lyapunov equation, a fundamental tool for stability analysis in physics and mathematics, and can be applied to other problems in the context of a linear response to parameter change~\cite{shitara_determining_2016,lostaglio_certifying_2020}.

\section{Acknowledgments}

G.~W. thanks Koenraad Audenaert and Dayou Yang for useful discussions that contributed to the development of the project. 
G.~W. acknowledges the Alexander von Humboldt Foundation for support under the Humboldt Research Fellowship. 
The authors acknowledge support of the EU project C-QuENS (grant no. 10113539) and by the state of Baden-Württemberg through bwHPC and the German Research Foundation (DFG) through grant no INST 40/575-1 FUGG (JUSTUS 2 cluster). 
M.M.R. acknowledges support of the National Science Center (NCN),
Poland, under projects No.~2020/38/E/ST3/00150. 
The project is developed together with an open-source repository~\cite{tn4qm}. 
The code is implemented using the open source library YASTN~\cite{yastn_scipost,yastn_codebase} that provides tensor network ansatz and basic algorithms. 

\appendix

\section{Quantum Fisher information}
\label{sec:convergence_general}

The Lyapunov equation for the SLD operator in Eq.~\eqref{eq:sld_def} is well-posed when the elements satisfy
\[\label{eq:sld_def_spectral}
\dT{\rhoParam{\param}}_{ij} = \frac{1}{2} [\sld]_{ij} \left(\lambda_i + \lambda_j\right) , 
\]
where $\hat O_{ij}=\bra{i}\hat O\ket{j}$ is a matrix element in the eigenbasis of the density matrix $\rhoParam{\param} = \sum_j\lambda_j\ket{j}\bra{j}$. 
When the density matrix is full rank, that is, $\forall\lambda_j: \quad\lambda_j>0$, the expression in Eq.~\eqref{eq:sld_def_spectral} is satisfied and all $[\sld]_{ij}$ elements are well defined. 
This is the case for any thermal state of finite temperature where we are sure to obtain eigenvalues exponentially decaying with the eigenvalue of the Hamiltonian. 
In this case the solution to the Eq.~\eqref{eq:sld_def} is given by the Lyapunov integral in Eq.~\eqref{eq:sld_integral}. 
The convergence of the SLD integral becomes evident when reformulated in its spectral representation
\[\label{eq:qfi_spectral_general}
\qfi(\argLim) = 2 \sum_{ij} |\dT{\rhoParam{\param}}_{ij}|^2 \int_0^\argLim d\arg \, e^{-(\lambda_i+\lambda_j)s} ,
\]
with exponential convergence guaranteed when the eigenvalues satisfy $\lambda_i+\lambda_j>0$. 

Potential issues with the definition and convergence of the SLD arise when the density matrix is singular, i.e. not full-rank. 
In particular, when $\lambda_i+\lambda_j=0$, that is, both eigenvalues are zero, the $\dT{\rhoParam{\param}}_{ij}$ on the left-hand side also has to be zero. 
The elements of the state derivative have a general form
\[\label{eq:drho_general}
\bra{i}\partial_\param{\rhoParam{\param}}\ket{j} = (\partial_\param\lambda_j) \delta_{ij} + \lambda_j \bra{i}{\partial_\param j}\rangle + \lambda_i \bra{\partial_\param i}{j}\rangle ,
\]
where $\partial_\param A\equiv\frac{\partial A}{\partial_\param}$ is the derivative over the estimated parameter $\param$. 
For the density matrix that is not full-rank, the consistency of the SLD in Eq.~\eqref{eq:sld_def_spectral} requires that the encoding produces zero velocities for every singular eigenvalue, i.e. $\forall \lambda_j=0: \quad \partial_\param\lambda_j=0$. 
If the SLD definition is consistent but the state is not full rank, the integral in Eq.~\eqref{eq:errqfi_spectral} converges as the divergent terms disappear and the singular elements of the SLD operator are arbitrary. 
If the SLD definition is ill-defined, pathological examples of QFI behavior that introduce discontinuities and discrepancies between the QFI and the Bures metric discussed in~\cite{safranek_discontinuities_2017,rezakhani_continuity_2019,seveso_discontinuity_2020} can occur. 
In such cases, the definition of the SLD must be regularized~\cite{liu_quantum_2014,rezakhani_continuity_2019}. 
The ill-defined cases of the encoding are beyond the scope of this paper. 

The convergence of the QFI integral in Eq.~\eqref{eq:qfi_spectral_general} depends on the spectrum of the density matrix and the state derivative in Eq.~\eqref{eq:drho_general}. 
The QFI error is
\[\label{eq:dqfi_spectral_general}
\qfi(\infty)-\qfi(\argLim) = 2 \sum_{ij}\frac{|\dT{\rhoParam{\param}}_{ij}|^2}{\lambda_i+\lambda_j} e^{-(\lambda_i+\lambda_j)\argLim} ,
\]
which is always positive, and the amplitude of the pair sum elements go down exponentially in the truncation cutoff $\argLim$ with the rate $1/(\lambda_i+\lambda_j)$. 
Since the state derivative $|\dT{\rhoParam{\param}}_{ij}|^2\sim(\lambda_i,\lambda_j)$ (up to the eigenstate deformation and $\partial_\param\lambda_j$ velocities), the decay of each term is weighted by $\frac{|\dT{\rhoParam{\param}}_{ij}|^2}{\lambda_i+\lambda_j}$, which is proportional to the eigenvalues $\lambda_i$ and $\lambda_j$. 
As a consequence, small-amplitude eigenvalues make only a minor contribution to the QFI, so truncating to a finite $\argLim$ yields a result that closely approximates the exact solution. 
However, this reasoning is valid only when the fraction of small eigenvalues is low. 
A discussion of the total error arising from small amplitudes can be found in Sec.~\eqref{sec:convergence_spectrum}.

\section{General bound on convergence error}
\label{sec:convergence_time}

Bounding the convergence error in Eq.~\eqref{eq:epsilon} that is defined with the QFI error in Eq.~\eqref{eq:errqfi_spectral} is a nontrivial task due to strong dependence on the spectrum of density matrix and the encoding operators. 
The scaling function for the error can be drastically different. 
For simplicity, let us consider the encoding has no diagonal terms. 
Then, a special case can be calculated for the pure state in which the convergence rate is
\[
\relErr(\argLim) = e^{-\argLim} ,
\]
implying exponential convergence regardless of the encoding operator and system size. 

A simple upper and lower bound of the relative error in Eq.~\eqref{eq:epsilon} can be calculated by noticing that $e^{-\argLim}\leq e^{-(\lambda_i+\lambda_j)\argLim} \leq e^{-\lambda_{\rm min} \argLim}$, where we used $\lambda_{\rm min}\leq\lambda_i+\lambda_j\leq1$
for any pair of eigenvalues $i\neq j$ and $\lambda_{\rm min}$ is smallest non-zero eigenvalue. 
Using this relation, we write the bound
\[
e^{-\argLim}\leq \relErr(\argLim) \leq  e^{-\lambda_{\rm min} \argLim},
\]
that converges exponentially to the upper bound of the integral $\argLim$. 
The smallest eigenvalue can scale badly with the system size, e.g. the seemingly trivial case of a product state with eigenvalues $\{1-p, p\}$ for each site has eigenvalues $\lambda_n = \alpha^n (1-p)^{N}$ $n\in[0,N]$ for $N$ subsystems, and the eigenvalues scale exponentially with $N$. 

Another bound can be calculated for the convergence error under a specific encoding. 
First, we notice that the convergence error in Eq.~\eqref{eq:errqfi_spectral} can be bounded by 
\[
\qfi(\infty) -\qfi(\argLim) \leq 2 \sum_{i,j\neq i} {|A_{ij}|^2 (\Lambda_i+\Lambda_j)} e^{-(\Lambda_i+\Lambda_j) \argLim} ,
\]
where we used that for non-negative eigenvalues $|\Lambda_i-\Lambda_j|\leq|\Lambda_i+\Lambda_j|$ and the function of eigenvalues can be bounded as follows 
\[
{(\Lambda_i+\Lambda_j)} e^{-(\Lambda_i+\Lambda_j) \argLim} \leq {\min}[({e\argLim})^{-1}, e^{-\lambda_{\rm min}\argLim}] ,
\]
where the first is an upper bound of the function $f(x)=xe^{-x\argLim}\leq({e\argLim})^{-1}$ and the second uses the slower rate $(\lambda_i+\lambda_j) e^{-(\lambda_i+\lambda_j) \argLim} \leq e^{-\lambda_{\rm min}\argLim}$, where $\lambda_{\rm min}={\rm min}_{i,j\neq0}[\lambda_i+\lambda_j]$ and we used that $\lambda_i+\lambda_j\leq 1$. 
Finally, we write the bound on the QFI error as follows
\[\label{eq:bound}
d\qfi(\argLim) \leq 2\, {\min}[({e\argLim})^{-1}, e^{-\lambda_{\rm min}\argLim}] \cdot {\sum_{i,j\neq i} |A_{ij}|^2} .
\]
We can additionally use $\sum_{i,j\neq i} |A_{ij}|^2\leq \tr(A^2)$ to make the formula independent of the ${\rhoParam{\param}}$ eigenbasis. 
A disadvantage of using $\tr(A^2)$ is the fact that it typically scales badly with the system size, for example, for $A = \sum_{j=1}^N \z{j}$ we have $\tr(A^2)=2^N N$, where $N$ is the system size. 
This problem can be reduced by returning to the bound in Eq.~\eqref{eq:bound} which is tighter but also harder to compute. 
Importantly, the upper bound for the worst-case scenario predicts that the error converges inversely proportional to the integration cutoff $\argLim$ or converges exponentially with the slowest decay rate $\lambda_{\rm min}\argLim$. 

\section{Low temperature approximation}
\label{sec:convergence_time_lowT}

In the low-temperature limit, the state can be approximated by the ground state and the first excited state. 
The approximation works well for the thermal states when $\beta\Delta\gg1$, where $\Delta$ is the energy gap. 
We develop the approximation for a low population of the excited state $p\ll1$. 
In the derivation, we assume that the population vector of the eigenstates is
\[\label{eq:lambda}
\vec\lambda=\left[\frac{1-p}{m}, \cdots, \frac{1-p}{m}, \frac{p}{n}, \cdots, \frac{p}{n}, 0, \cdots, 0\right], 
\]
where a population of the ground state is $1-p$ and it is $m$-fold degenerate and $p$ is a population of the excited level and it is $n$-fold degenerate.

The convergence error in Eq.~\eqref{eq:epsilon} for the approximation reads
\[\label{eq:eps_low_temp_full}
\begin{split}
\relErr(\argLim) = \frac{1}{\qfi(\infty)} 2\bigg(&c_1 \frac{1-p}{m}e^{-\frac{1-p}{m}\argLim} + c_2 \frac{p}{n}e^{-\frac{p}{n}\argLim} \\
&+c_3 \frac{(\frac{1-p}{m}-\frac{p}{n})^2}{\frac{1-p}{m}+\frac{p}{n}}e^{-(\frac{1-p}{m}+\frac{p}{n})\argLim}\bigg), 
\end{split}
\]
where the constants
\begin{align*}
c_1 &=  2\sum\nolimits_{\bar m:\lambda_{\bar m}=(1-p)/m}\sum\nolimits_{j:\lambda_j=0} |A_{ij}|^2,\\
c_2 &=  2\sum\nolimits_{\bar n:\lambda_{\bar n}=p/n}\sum\nolimits_{j:\lambda_j=0} |A_{ij}|^2,\\
c_3 &= 2\sum\nolimits_{\bar m:\lambda_{\bar m}=(1-p)/m}\sum\nolimits_{\bar n:\lambda_{\bar n}=p/n} |A_{ij}|^2,
\end{align*}
that are the sums of elements of the encoding operator. 
We upper bound the expression by noticing that $\frac{(x-y)^2}{x+y}e^{-(x+y)\argLim}\leq|x-y|e^{-|x-y|\argLim}$ where we have $|x-y|>0$ and obtain the expression 
\[\label{eq:eps_low_temp_bound}
\begin{split}
\relErr(\argLim) \leq \frac{1}{\qfi(\infty)} 2\bigg(&c_1 \frac{1-p}{m}e^{-\frac{1-p}{m}\argLim} + c_2 \frac{p}{n}e^{-\frac{p}{n}\argLim} \\
&+\frac{c_3}{m}\left|1-p\frac{m+n}{n}\right|e^{-|1-p\frac{m+n}{n}|\argLim/m}\bigg) .
\end{split}
\]
In the next step, we apply the small $p$ limit by assuming that $p\ll n/(m+n)$ which is still a useful limit when the GS degeneracy $m$ is moderate in comparison to ES degeneracy $n$. 
In this regime, the bound reads
\[\label{eq:eps_low_temp_approx}
\begin{split}
\relErr(\argLim) \lesssim \frac{1}{\qfi(\infty)} 2\bigg(&\frac{c_1+c_3}{m}e^{-\argLim/m} + p \frac{c_2}{n}e^{-\frac{p}{n}\argLim}\bigg) .
\end{split}
\]
Finally, we linearize the formula by noticing that $p\, e^{-{p\argLim}/{n}}\leq p$ for small $p$ 
\[\label{eq:eps_low_temp_approx2}
\begin{split}
\relErr(\argLim) \lesssim \frac{1}{\qfi(\infty)} 2\bigg(&\frac{c_1+c_3}{m}e^{-\argLim/m} + p \frac{c_2}{n}\bigg) .
\end{split}
\]

In order to make a connection between the convergence error and the entropy, we use the relation
\[\label{eq:p2EE}
\sum_j \lambda_j (1-\lambda_j) \leq {S(\vec \lambda)}/(2\log2) ,
\]
where $S(\vec \lambda)$ is von Neumann entropy. 
See also Ref.~\footnote{
The bound can be derived by noticing that~\cite{chu_thermodynamic_2022}
\[
f(p) = -p\log p +  2 p^2\log2 ,
\]
for which $f(p_1+p_2) \leq f(p_1)+f(p_2)$. 
For the probability distribution with normalization $\sum_jp_j=1$ and the entropy defined as $S(p) = -\sum_j p_j\log p_j$ we find the relation in Eq.~\eqref{eq:p2EE}. 
}. 
The bound is a good approximation for low entropy states. 
The left-hand side of the relation in Eq.~\eqref{eq:p2EE} reads
\[
\sum_j \lambda_j (1-\lambda_j) = m \frac{1-p}{m} \left(1-\frac{1-p}{m}\right) + n \frac{p}{n} \left(1-\frac{p}{n}\right) ,
\]
which reduces to 
\[\label{eq:LLm1}
\sum_j \lambda_j (1-\lambda_j) \approx \frac{m-1}{m} + \frac{2}{m}p ,
\]
when we used that $p\ll n/(m+n)$ which is consistent with the derivation of the convergence error so far. 
Plugging it into the relation in Eq.~\ref{eq:p2EE} we obtain
\[\label{eq:p2S}
p \lesssim {S(\vec \lambda)} \frac{m}{4\log2} - \frac{m-1}{2} .
\]

In the end, we write the bound on the convergence error in Eq.~\eqref{eq:eps_low_temp_final} by using Eq.~\eqref{eq:p2S} and the approximate QFI $\qfi(\infty)\gtrsim 2 (\frac{c_1+c_3}{m})$ where we additionally dropped the term proportional to $c_2$. 
Using the approximation, we obtain the bound
\[\label{eq:eps_low_temp_final}
\relErr(\argLim) \lesssim e^{-\argLim/m} + \frac{c}{2} \frac{m}{n} \left(S(\vec \lambda)\frac{m}{2\log2}-m+1\right) ,
\]
where $c=\frac{c_2}{c_1+c_3}$. 
For non-degenerate spectrum $m=n=1$ reads
\[
\relErr(\argLim) \lesssim e^{-\argLim} + \frac{c}{2\log4} S(\vec \lambda) .
\]
The comparison of numerical data and the bounds derived in this chapter is presented in Fig.~\ref{fig:fig3} in the main text. 

\section{Estimate convergence from the spectrum}
\label{sec:convergence_spectrum}

The convergence error of the QFI integrand in Eq.~\eqref{eq:errqfi_spectral} is generally difficult to estimate. 
Simple bounds have been presented in Sec.~\ref{sec:convergence}, these, however, are not always useful or are limited to specific cases. 
In this section, we discuss the estimate of the convergence error based on a partial information about the eigenspectrum of the density matrix. 
Partial information about the spectrum can be obtained numerically for the state using standard Krylov methods where we find the first few dominant eigenvalues. 

We assume that the spectrum of the density matrix is split into $\Lambda$, which is a set of known eigenvalues, and $\tilde\Lambda$, such that $\forall i,j: \quad\Lambda_i\geq\tilde\Lambda_j$, which is a complement to the known set. 
We note that the error function can be divided into three terms
\[
d\qfi(\Lambda,\Lambda) + 2 \cdot d\qfi(\Lambda,{\tilde\Lambda}) + d\qfi({\tilde\Lambda},{\tilde\Lambda}) ,
\]
where
\[\label{eq:dqfi}
d\qfi(\vec p,\vec q) = 2 \sum_{i\in\vec p} \sum_{j\in\vec q} \frac{|A_{ij}|^2 (p_i-q_j)^2}{p_i+q_j} e^{-(p_i+q_j) \argLim} .
\]
The contribution of the known set, that is, $d\qfi(\Lambda,\Lambda)$, can be calculated explicitly using the eigenvalues and eigenstates that we have. 
On the other hand, the remaining terms include the contribution of the unknown set $\tilde\Lambda$ and cannot be computed. 
We notice that although these take smaller amplitudes of eigenvalues, they can be meaningful due to number of summed terms. 
In particular, the term $ d\qfi(\Lambda,{\tilde\Lambda})$ sums over $n_\Lambda n_{\tilde\Lambda}$ elements and $d\qfi(\tilde\Lambda,{\tilde\Lambda})$ sums over $n_{\tilde\Lambda} n_{\tilde\Lambda}$ elements, where $n_\Lambda=\dim\Lambda$ and $n_{\tilde\Lambda}=\dim\tilde\Lambda$. 

If we know only a few dominant eigenpairs of the density matrix, i.e. $n_\Lambda$ is small, the contribution $d\qfi(\tilde\Lambda,{\tilde\Lambda})$ accumulate the majority of terms. 
To simplify the unknown contribution we invoke the bound
\[\label{eq:dqfi_minus_to_plus}
d\qfi(\vec p,\vec q) \leq 2 \sum_{i\in\vec p} \sum_{j\in\vec q\setminus\vec p} {|A_{ij}|^2 (p_i+q_j)} e^{-(p_i+q_j) \argLim} ,
\]
where we used $|p_i-p_j|\leq|p_i+p_j|$ for real numbers and diagonal elements are excluded from the sum, i.e. for $\vec p=\vec q$ we omit $i=j$ elements. 
We notice that the error in Eq.~\eqref{eq:dqfi_minus_to_plus} is Shur concave~\cite{schur1923uber,ostrowski_sur_1984} when $p_i+q_j\leq 1/\argLim$. 
We use this property to bound $d\qfi(\tilde\Lambda,{\tilde\Lambda})$ under the assumption that
\[
\argLim\leq\frac{1}{2\Lambda_{\rm min}} ,
\]
that coincides with the regime where $d\qfi(\Lambda,\Lambda)$ is still relevant (the slowest decay rate is $1/2\Lambda_{\rm min}$. 
The bound on the error contribution reads 
\[\label{eq:dqfi_tilde_Lambda_bound}
d\qfi(\tilde\Lambda,{\tilde\Lambda}) \leq 4 c_{\tilde\Lambda} {p_{\tilde\Lambda}}(n_{\tilde\Lambda}-1) e^{-{2p_{\tilde\Lambda}}X/n_{\tilde\Lambda}} ,
\]
where $c_{\tilde\Lambda}=\max_{i\in\tilde\Lambda, j\in\tilde\Lambda}|A_{ij}|^2$ and $p_{\tilde\Lambda}=\tr[\tilde\Lambda]=1-\tr[\Lambda]$ is a sum of unknown elements and we used the flat distribution with each element being ${p_{\tilde\Lambda}}/{n_{\tilde\Lambda}}$. 
The formula in Eq.~\eqref{eq:dqfi_tilde_Lambda_bound} indicates that the error of neglected terms increases with the population of the unknown set $p_{\tilde\Lambda}$ and the dimension $n_{\tilde\Lambda}$ over which it is distributed. 

The remaining error term mixing contributions $\Lambda$ and $\tilde\Lambda$ can be estimated by the following formula
\[
d\qfi(\Lambda,\tilde\Lambda) \leq 2 \sum_{i\in\Lambda} \sum_{j\in\tilde\Lambda} {|A_{ij}|^2 (\Lambda_i+\Lambda_{\rm min})} e^{-\Lambda_i \argLim} ,
\]
where we used $e^{-(\Lambda_i+\tilde\Lambda_j) \argLim}\leq e^{-\Lambda_i \argLim}$ and $\Lambda_i+\tilde\Lambda_j\leq\Lambda_i+\Lambda_{\rm min}$. 

Finally, it is important to highlight that the bound we establish relies on the constants proportional to $A_{ij}$ that require prior knowledge of the eigenvectors of the density matrix for their computation.  
Consequently, when only a fixed set $\Lambda$ and its associated eigenvectors are available, the bounds only provide information about the functional scaling of the QFI error, rather than giving explicit error values.

\section{Unitary encoding}
\label{sec:operator_vs_general}

In the main text, we introduced two possible implementations for unitary encoding: the standard SLD integration in Sec.~\ref{sec:implementation} and integration of the encoding operator in Sec.~\ref{sec:unitary}. 
These two approaches offer different strategies for handling the encoding process. 

\begin{figure}[b!]
    \centering
    \includegraphics[width=\columnwidth]{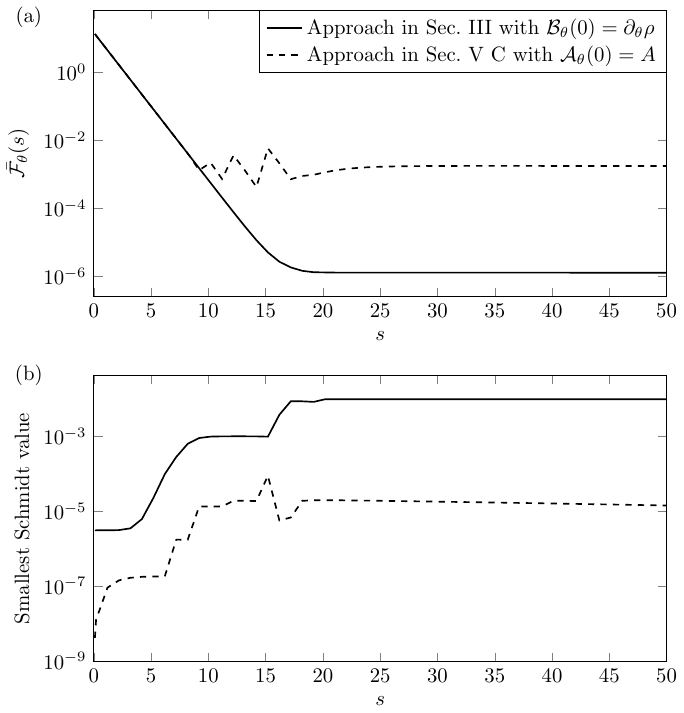}
    \caption{{\bf Errors of integration schemes.}
    Comparison of the two implementations for the unitary encoding. 
    The panels show (a) the QFI integrand and (b) the smallest Schmidt value for the integrand. 
    Data are complementary to Fig.~\ref{fig:fig4} and are obtained for the QFI integrand $\dqfi(\arg)=2\tr[\dT{\rhoParam{\param}} \B(\arg)]$ where the $\B(\arg)$ as in Sec.~\ref{sec:implementation} (solid) or Sec.~\ref{sec:unitary} (dashed). 
    Despite the error for $\A$-based integration, its error on the QFI integral is $0.48\%$ of the $\qfi(50)$ value.  
    }
    \label{fig:fig6}
\end{figure}

The encoding integration in Sec.~\ref{sec:unitary} has a benefit of evolving the operator $A$ directly which typically has smaller bond dimension than $\dT{\rhoParam{\param}}$ operator. 
However, it is important to remember that calculating the QFI for the operator is costly and requires a contraction $\qfi(\argLim)= 2\, \tr([A,{\rhoParam{\param}}]\,  [\K(\argLim),{\rhoParam{\param}}])$, which required a contraction involving a product between four MPO operators. 
For this reason, it is discouraged operation. 
The approach can still be useful for obtaining the SLD operator when the QFI is not evaluated often. 

The compression of the MPO operators is necessary to control the computational cost. 
In Fig.~\ref{fig:fig6}, we see non-trivial impact of truncation for the QFI integrand. 
In the figure, the approach based on $\A(\arg)$ integrand gives larger error than standard $\B(\arg)$, see Fig.~\ref{fig:fig6}a, despite truncating only the low Schmidt values, see Fig.~\ref{fig:fig6}b. 
A possible explanation of such behavior is the property of the compression. 
The MPO compression scheme is optimal with respect to the operator 2-norm but may not be optimal for more complicated observables. 
Therefore, for the QFI integrand $\dqfi(\arg)= ||\B(\arg/2)||_2^2$ the truncation of $\B(\arg/2)$ is optimal for $\dqfi(\arg)$. 
For the approach based on operator $A(\arg)$ it is not the case. 
The situation may be similar for $\dqfi(\arg)=2\tr[\dT{\rhoParam{\param}} \B(\arg)]$ as observed in Fig.~\ref{fig:fig6}. 
This observation suggest caution when proceeding with the integration scheme in Sec.~\ref{sec:unitary}. 

\section{Cost of tensor network implementation}
\label{sec:computational_cost}

The integration method is based on the TDVP update
\[\label{eq:update}
\Omega(\arg+\Darg) = e^{-{\rhoParam{\param}} \Darg}\Omega(\arg)e^{-{\rhoParam{\param}} \Darg} , 
\]
where $\Omega(\arg)$ is $\B(\arg) = e^{-{\rhoParam{\param}} \arg}\dT{\rhoParam{\param}} e^{-{\rhoParam{\param}} \arg}$ in standard version in Sec.~\ref{sec:implementation} or $\A(\arg) = e^{-{\rhoParam{\param}} \arg}Ae^{-{\rhoParam{\param}} \arg}$ for integration of the encoding in Sec.~\ref{sec:unitary}. 
In this paper, we used the implementation in the YASTN package~\cite{yastn_scipost,yastn_codebase} where the integrand can be propagated by the sum of bra and ket operators. 
It allows one to propagate the integrand without explicitly constructing the propagator $\rhoParam{\param}\otimes \mathbbm 1 + \mathbbm 1\otimes\rhoParam{\param}^T$ generating the exponent acting on $bra$ and $ket$ dimensions together. 

The cost of a single update in Eq.~\eqref{eq:update} scales as $\mathcal{O}(d^2D_\Omega^2D_{\rhoParam{\param}} \, {\rm max}[dD_{\rhoParam{\param}}, \, D_\Omega])$ which is $\mathcal{O}([D_\Omega^2, D_\Omega^3])$ in terms of the $\Omega$ bond dimension, and $\mathcal{O}([D_{\rhoParam{\param}}^2, D_{\rhoParam{\param}}])$ in terms of the ${\rhoParam{\param}}$ bond dimension ($D_\Omega<dD_{\rhoParam{\param}}$ for the former and $D_\Omega>dD_{\rhoParam{\param}}$ for the latter). 
Notice that the integration scheme is independent of the SLD bond dimension. 
The SLD is calculated during the integration, and there the cost is related to MPO-MPO addition between the $\sld$ and the full SLD integrand $\B(\arg)$. 
During the addition, the bond dimension can increase at most to $D_\sld(\argLim)+D_\B(\argLim)$. 
To ensure a fast reduction of the SLD bond dimension after the update, we can use probabilistic low-rank factorization ~\cite{tamascelli_improved_2015,kohn_probabilistic_2018} which is useful for large dimensionality reduction. 

\begin{figure}[t!]
    \centering
    \includegraphics[width=\columnwidth]{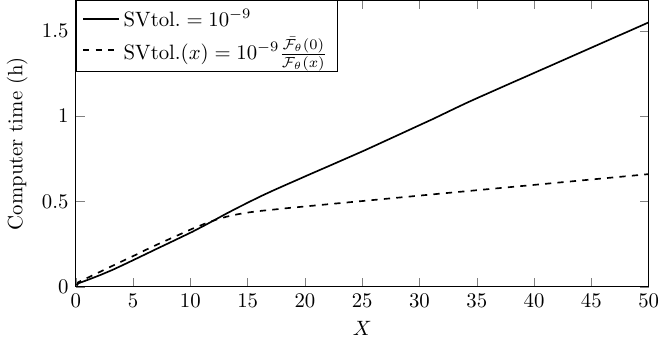}
    \caption{
    {\bf Dynamical compression.}
    Comparison of the computational time for two scenarios: a fixed truncation threshold of $\SVtol=10^{-9}$ (solid) and a dynamic compression approach where $\SVtol(\arg)=10^{-9}\frac{\dqfi(0)}{\dqfi(\arg)}$ (dashed).
    The dynamic compression method employs more aggressive truncation when the integrand amplitude is low, thereby reducing computational costs. 
    Data obtained for $N=16$, $\tf=2J$, $\beta=4J^{-1}$, bond dimensions $D_{\rhoParam{\param}}\leq32$, $D_\B\leq64$ and integration step $\Darg=0.1$. 
    Computer time obtained for $4$ CPU cores. 
    }
    \label{fig:fig7}
\end{figure}

The bond dimension of the integrand is also controlled by MPO compression. 
Motivated by the decay of the integral, we suggest an alternative truncation scheme in which the truncation cutoff is proportional to the amplitude of the integrand. 
Using this strategy, we apply stronger truncation when the amplitude drops down significantly. 
By allowing for more aggressive truncation, we can focus computational resources on the most significant parts of the integrand. 
We choose the truncation cutoff to be relative to the initial integrand amplitude so that the truncation cutoff
\[\label{eq:SVs}
\SVtol(\arg) = \tilde\relErr|\B(0)|/|\B(\arg/2)| ,
\]
where $\tilde\relErr$ is a control parameter and $\SVtol(\arg)$ is now time dependent. 
Figure~\ref{fig:fig7} illustrates the variations in computational cost between the static and dynamic compression schemes. 
The implementation of an adaptive truncation cutoff demonstrates a substantial reduction in computational expenses. 
The change in plot monotonicity in Fig.~\ref{fig:fig7} coincides with a change in the dominant contribution to the SLD integrand. 

\section{Variational approach}
\label{sec:varia}

The variational method introduced in Ref.~\cite{macieszczak_quantum_2013,chabuda_tensor-network_2020} relies on the QFI figure of merit
\[\label{eq:qfi_varia}
\qfi = 2\, \tr{[\dT{\rhoParam{\param}}\sld]} - \tr{[{\rhoParam{\param}}\sld^2]} ,
\]
where $\sld$ is the symmetric logarithmic derivative, ${\rhoParam{\param}}$ is the state density matrix and $\dT{\rhoParam{\param}}$ is the state derivative. 
The definition of QFI in Eq.~\eqref{eq:qfi_varia} is general and exactly equivalent to the expression in Eq.~\eqref{eq:qfi}. 
In the variational approach, we search for the extremum of the QFI in Eq.~\eqref{eq:qfi_varia} over the $\sld$ operators. 
Notice that the extremum, i.e. $0 = 2 \dT{\rhoParam{\param}} - \sld{\rhoParam{\param}} - {\rhoParam{\param}}\sld$, corresponds to the SLD definition in Eq.~\eqref{eq:sld_integral} which ensures that we get a correct solution. 

In the numerical implementation of the variation procedure, we perform local optimizations of the SLD operator. 
An update of the SLD tensor of the MPO representation translates to a linear problem of the form $Ax=b$ where $A$ derives from an effective operator calculated for the term $\tr{[{\rhoParam{\param}}\sld^2]}$ and $b$ from an effective operator of $2\, \tr{[\dT{\rhoParam{\param}}\sld]}$. 
The cost of calculating the $A$ and $b$ operators scales as $\mathcal{O}(d^2D_\sld^2D_{{\rhoParam{\param}}}{\rm max}[dD_{{\rhoParam{\param}}}, \, D_\sld])$ for the first term and $\mathcal{O}(d^2D_\sld D_{\dT{\rhoParam{\param}}}{\rm max}[D_{\dT{\rhoParam{\param}}}, \, D_\sld])$ for the second term. 
The combined scaling of the contraction is $\mathcal{O}(d^2D_\sld{\rm max}[dD_{{\rhoParam{\param}}}^2D_\sld, \, D_{{\rhoParam{\param}}}D_\sld^2, \, D_{\dT{\rhoParam{\param}}}^2, \, D_{\dT{\rhoParam{\param}}}D_\sld])$. 
For local encoding, i.e., encoding operator $A$ is a sum of local operators, thus the bond dimension is $D_A=2$, we can assume that $D_{\dT{\rhoParam{\param}}}\leq 4D_{{\rhoParam{\param}}}$, which translates to the scaling $\mathcal{O}(d^2D_\sld^2D_{{\rhoParam{\param}}}{\rm max}[dD_{{\rhoParam{\param}}}, \, D_\sld])$ assuming that $D_\sld\geq16$. 
In conclusion , the cost scales $\mathcal{O}([D_\sld^2, D_\sld^3])$ in the $\sld$ bond dimension and $\mathcal{O}([D_{\rhoParam{\param}}^2, D_{\rhoParam{\param}}])$ in the bond dimension of ${\rhoParam{\param}}$ ($D_\sld<dD_{\rhoParam{\param}}$ for the former and $D_\sld>dD_{\rhoParam{\param}}$ for the latter). 

The linear problem for local update can be solved with the Moore-Penrose pseudoinverse. 
While other approaches exit, the pseudoinverse has a benefit of controlling singular values that can be truncated from the solution which stabilizes the solution. 
In the original implementation in Ref.~\cite{chabuda_tnqmetro_2022}, the cost of pseudoinverse scales as $\mathcal{O}(d^6D_\sld^6)$ which is cubic with $d^2D_\sld^2$. 
This makes up a significant contribution to the computational cost with increasing bond dimension of the SLD operator. 
Additionally, the original implementation does not allow dynamical updates to the variational method, may have a problem with stability due to the numerical inverse, the current approach does not maintain the canonical form of the MPO, resulting in a less compact MPO representation, and cannot perform local updates on the MPO ansatz, necessitating a complete restart of the simulation whenever modifications to the SLD ansatz are required. 

The scaling can be greatly improved by exploiting the Krylov methods. 
There, the pseudoinverse can be solved on an auxiliary problem defined in the Krylov basis~\cite{Lloyd_numerical_1997}. 
There we construct the auxiliary problem defined in the basis of Krylov vectors. 
The pseudoinverse if calculated on the level of reduced problem which has much smaller dimension that the full problem. 
The main cost of the Krylov expansion was reduced to iterative application of the effective operator $A$ to expand the Krylov basis as explained earlier in the text. 
The Krylov approach to the variational method was implemented in the TN4QM~\cite{tn4qm} repository and can be accessed by the reader. 
The comparison of the numerical cost of a single update is presented in Fig.~\ref{fig:fig5}.

\section{Adaptive integration step}
\label{sec:adaptive}

Numerical integration with a fixed integration step leads to a numerical error that depends on the functional form of the integral. 
In particular, it may introduce a large error when the decay rate is faster than the step, on the other hand, it may be unnecessarily small for a slowly varying integrand. 
To control the integration error, we develop an adaptive scheme in which the step size is dynamically adjusted based on the QFI integrand. 
By setting an appropriate tolerance parameter for this adaptive method, we can effectively maintain the integration error below a specified threshold. 

\begin{figure}[t!]
    \centering
    \includegraphics[width=\columnwidth]{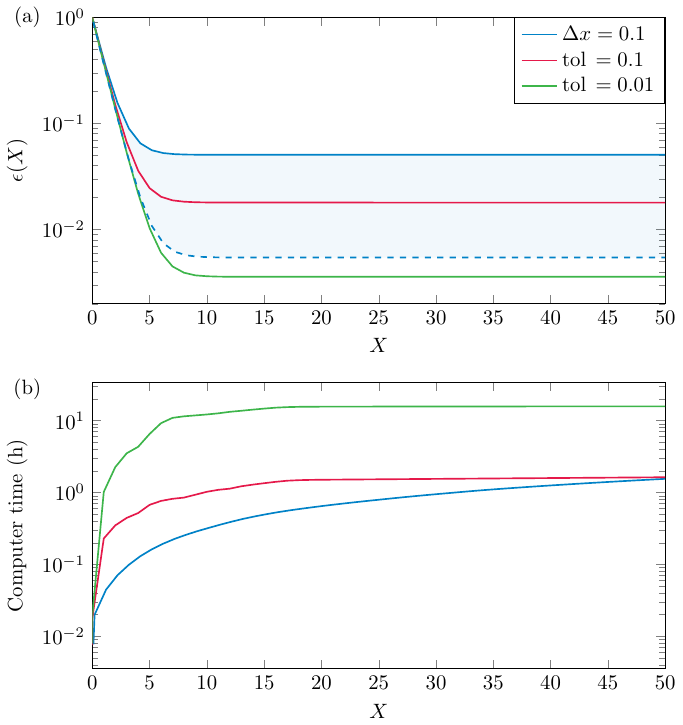}
    \caption{{\bf Adaptive integration step.}
    (a) A comparison between the integration with a fixed step $\Darg=0.1$ (blue) and the adaptive scheme with $\adtol=0.1$ (red) and $\adtol=0.01$ (green). 
    The bound on the integration error for the fixed integration step in Eq.~\eqref{eq:e_bound} is primarily accumulated in the beginning of simulation where the integrand decays quickly (dashed blue line and shaded area). 
    (b) Numerical cost of the simulation measured by the computational time. 
    For the adaptive scheme, necessity to make smaller steps when the integrand changes quickly may increase the computational cost. 
    Data obtained for $N=16$, $\tf=2J$, $\beta=4J^{-1}$, bond dimensions $D_{\rhoParam{\param}}\leq32$ and $D_\B\leq64$. 
    Computer time obtained for $4$ CPU cores. 
    }
    \label{fig:fig8}
\end{figure}

In the procedure, we choose a discrete step $\Darg$ such that the following condition is met
\[\label{eq:adaptive_tol}
\frac{\dqfi(\arg) - \dqfi(\arg+\Darg)}{\dqfi(\arg+\Darg)} \leq \adtol , 
\]
where $\adtol$ is the tolerance parameter for the adaptive scheme. 
For the quadrature rule, the error of a single integration step is
\[\label{eq:error_basic}
e(\arg, \arg+\Darg) = \int_\arg^{\arg+\Darg} d\arg' \dqfi(\arg') - \Darg_j {\dqfi(\arg+\Darg)} ,
\]
where $\qfi^{\rm num}(\argLim) = \sum_j \Darg_j {\dqfi(\arg+\Darg)}$ is the numerical approximation of the integral at $\argLim$ and $\sum_j \Darg_j=\argLim$. 
The choice of quadrature rule is such that the numerical approximation gives a lower bound on the exact QFI. 
We do so to avoid overestimating the ultimate precision in Eq.~\eqref{eq:cramerrao}. 
The error of the discrete approximation is
\[\label{eq:e_bound}
0\leq e(\arg, \arg+\Darg)\leq \frac{1}{2} \Darg_j \, \adtol \dqfi(\arg+\Darg).
\]
where we used $\dqfi(\arg) - \dqfi(\arg+\Darg)\leq \adtol \dqfi(\arg+\Darg)$ from the condition in Eq.~\eqref{eq:adaptive_tol}. 
Summing the errors in all steps up to the cutoff point $\argLim$ we obtain the integrated error
\[\label{eq:error_adaptive}
E(\argLim) = \sum_j e(\arg_j, \arg_j+\Darg_j) \leq \frac{\adtol}{2} \qfi^{\rm num}(\argLim)
\]
where the total error on the QFI is $E(\argLim) = \qfi(\argLim) - \qfi^{\rm num}(\argLim)$. 
Notice that in this case $\qfi^{\rm num}(\argLim) < \qfi(\argLim)$. 
The error can be related to the relative error in Eq.~\eqref{eq:epsilon}
\[\label{eq:epsilon_adaptive}
\relErr_\adtol(\argLim) \leq \frac{\adtol}{2} \frac{\qfi^{\rm num}(\argLim)}{\qfi(\infty)}
\]
which for ${\qfi^{\rm num}(\argLim)}\leq{\qfi(\infty)}$ reduces to a less tight bound $\relErr_\adtol(\argLim) \leq \frac{\adtol}{2}$. 
Similar consideration can be made for the trapezoid rule, and the error is the same but an opposite sign. 

Figure~\ref{fig:fig8} illustrates a comparison between the integration with a fixed integration step and the adaptive approach. 
The adaptive approach allows to control the integration error, however, it may require a larger number of steps in regions where the function exhibits rapid changes. 
This trade-off between accuracy and computational efficiency is a key consideration when using the adaptive method.

\end{document}